\documentclass[useAMS,usenatbib]{mn2e}
\usepackage{bm}
\usepackage{amsfonts}
\usepackage{amsmath}
\usepackage{graphicx}
\usepackage{subfigure}

\title[Dependence of AGN spectral index on luminosity]{Is the dependence of spectral index on luminosity real in optically selected AGN samples?}
\author[S. M. Tang, S. N. Zhang and P. F. Hopkins]{Su Min Tang $^{1,\ 5}$\thanks{E-mail:
stang@cfa.harvard.edu (SMT); zhangsn@tsinghua.edu.cn
(SNZ)}, Shuang Nan Zhang $^{1,\ 2,\ 3,\ 4}$ and Philip F. Hopkins $^{5}$\\
$^{1}$Department of Physics and Center for Astrophysics,
Tsinghua University, Beijing 100084, China \\
$^{2}$Department of Physics, University of Alabama in Huntsville,
Optics Building 201C, Huntsville, AL 35899 \\
$^{3}$Space Science Laboratory, NASA
Marshall Space Flight Center, SD50, Huntsville, AL 35812 \\
$^{4}$Institute of High Energy Physics, Chinese Academy of Sciences,
P.O. Box 918-3, Beijing 100039, China \\
$^{5}$Harvard-Smithsonian Center for Astrophysics, 60 Garden Street,
Cambridge, MA 02138}

\begin{document}

\date{Accepted date. Received  date}

\pagerange{\pageref{firstpage}--\pageref{lastpage}} \pubyear{2002}

\maketitle

\label{firstpage}

\begin{abstract}
We critically examine the dependence of spectral index on luminosity
in optically selected AGN samples. An analysis of optically selected
high-redshift quasars showed an anti-correlation of $\alpha_{OX}$,
the spectral index between the rest-frame 2500 \AA\ and 2 keV, with
optical luminosity (Miyaji et al. 2006). We examine this
relationship by means of Monte Carlo simulations and conclude that a
constant $\alpha_{OX}$ independent of optical luminosity is still
consistent with this high-z sample. We further find that that
contributions of large dispersions and narrow range of optical
luminosity are most important for the apparent, yet artificial,
$\alpha_{OX} - l_o$ correlation reported. We also examine another,
but more complete low-z optical selected AGN sub-sample from Steffen
et al. (2006), and our analysis shows that a constant $\alpha_{OX}$
independent of optical luminosity is also consistent with the data.
By comparing X-ray and optical luminosity functions, we find that a
luminosity independent $\alpha_{OX}$ is in fact more preferred than
the luminosity dependent $\alpha_{OX}$ model. We also discuss the
selection effects caused by flux limits, which might systematically
bias the $l_x - l_o$ relation and cause discrepancy in optically
selected and X-ray selected AGN samples. To correctly establish a
dependence of $\alpha_{OX}$ of AGNs on their luminosity, a larger
and more complete sample is needed and consequences of luminosity
dispersions and selection effects in flux limited samples must be
taken into account properly.
\end{abstract}


\section{Introduction}
The dependence of the spectral index $\alpha_{OX}$ of active
galactic nuclei (AGN) on redshift and luminosity has important
astrophysical implications on AGN evolution and thus has been
studied for many years (e.g. Avni $\&$ Tananbaum 1982; Wilkes et al.
1994; Green et al. 1995; Bechtold et al. 2003; Vignali et al. 2003a;
Strateva et al. 2005; Steffen et al. 2006; Hopkins et al. 2007;
Kelly et al. 2007). $\alpha_{OX}$ is defined as
\begin{equation}
\alpha_{OX}=\frac{\log(f_{2\ \rm keV}/f_{2500\ \rm \AA})}
{\log(\nu_{2\ \rm keV}/\nu_{2500\ \rm \AA})},
\end{equation}
where $f_{2\ \rm keV}$ and $f_{2500\ \rm \AA}$ are the rest-frame flux densities at 2
keV and 2500 $\rm \AA$, respectively. Dependence of $\alpha_{OX}$ on redshift means
evolution of the accretion process in cosmic time. Most studies have concluded that
there is no evidence for a dependence of $\alpha_{OX}$ on redshift (e.g. Avni $\&$
Tananbaum 1982; Strateva et al. 2005), although some studies found that $\alpha_{OX}$
is correlated with redshift (Bechtold et al. 2003; Kelley et al. 2007). Dependence of
$\alpha_{OX}$ on luminosity means a non-linear relationship between X-ray and optical
luminosity ($L_X \propto L_O^e$, $e\neq1$), which provides insight into the radiation
mechanism. An anti-correlation between $\alpha_{OX}$ and the optical luminosity has
been found by many authors in optically selected AGNs with follow-up X-ray measurements
at some different epoch, which means that these AGNs span a larger range in optical
luminosity than in X-ray luminosity (e.g. Vignali et al. 2003a; Strateva et al. 2005;
Miyaji et al. 2006, hereafter M06; Steffen et al. 2006, hereafter S06). Meanwhile,
whether $\alpha_{OX}$ depends on luminosity in X-ray selected AGN samples remains
unknown (Hasinger 2004; Frank et al. 2007).

As pointed out by Yuan et al. (1998), one of the problems in such
studies is that an apparent, yet artificial correlation between
$\alpha_{OX}$ and optical luminosity can be caused by dispersions in
the optical luminosity. In section 2, we present analysis of the
relationship between $\alpha_{OX}$ and optical/X-ray luminosity
using data presented in M06, in which we find that dispersions in
luminosity can be entirely responsible for the claimed dependence.
We also discuss the determining factor for this behavior.

Another problem is the degeneracy between redshift and luminosity in
flux-limited samples, where redshift and luminosity are strongly
correlated. In section 3, we examine a sub-sample from S06
containing 187 AGNs, which more completely fills the redshift and
optical luminosity plane and thus is less affected by such
degeneracy. In section 4, we compare the optical quasar luminosity
function from Richards et al. (2006) with X-ray quasar luminosity
function from Barger et al. (2005) in different $\alpha_{OX}$
models. In section 5, we discuss the selection effects in flux
limited samples and the consequently discrepancy in optically
selected and X-ray selected samples. Discussion and conclusions are
presented in section 6.

We mostly use the logarithms of luminosities and denote them as
$l_x=\log L_{2\ \rm keV}$ and $l_o=\log L_{2500}$, then
$\alpha_{OX}=0.3838(l_x-l_o)$, where $L_{2\ \rm keV}$ is the 2 keV
monochromatic luminosity and $L_{2500}$ is the 2500 $\rm \AA$
monochromatic luminosity in units of erg s$^{-1}$ Hz$^{-1}$. We
adopt the currently favored cosmology model with $H_0$=70 km
s$^{-1}$ Mpc$^{-1}$, $\Omega_M=0.3$, and $\Omega_{\Lambda}=0.7$
(e.g. Spergel et al. 2007).

\section{Analysis of Miyaji et al. (2006) Sample}

\subsection{Data Analysis}
The sample we use in this section is consisted of 61 high-redshift
($z>2.9$) quasars in Figure 2 of M06 (Miyaji et al. 2006; Vignali et
al. 2003b, 2005), excluding the three quasars from archival $\it
Chandra$ data in Table 3 in M06 which might be biased toward higher
X-ray fluxes. Only six of them have no X-ray detection. M06 found a
correlation of $\alpha_{OX}$ with optical luminosity for this high-z
sample, while they kept the discussion on the $l_o - \alpha_{OX}$
relation open because of possible optical selection effects for
variable AGNs which preferentially pick up the optically brighter
phases.

To illustrate how dispersions produce artificial correlations in
this sample, we carry out two independent analysis. The first one is
linear regression of $\alpha_{OX}$, $l_o$ and $l_x$ in observed
data. Without further description, we perform linear regression
using methods as follows throughout the paper:
\begin{enumerate}
  \item For the $\alpha_{OX} - l_o$ correlation, we use the EM
algorithm in ASURV (Isobe, Feigelson, \& Nelson 1986) to derive
linear regression parameters, including X-ray undetected quasars;
  \item The EM linear regression algorithm in ASURV is based on the
traditional ordinary least-squares method which minimizes the
residuals of the dependent variable (OLS(Y$\mid$X)). However, for
$l_x - l_o$ correlation, both variables are observed and a different
result can be obtained if residuals of the independent variable are
instead minimized (e.g. S06). Following S06, we perform linear
regression with ASURV using EM algorithm, treating $l_x$ as the
dependent variables (ILS(Y$\mid$X)) and treating $l_o$ as the
dependent variables (ILS(X$\mid$Y)), then use the equations given by
Isobe et al. (1990) to calculate the bisector of the two regression
lines.
  \item For the $\alpha_{OX} - l_x$, where both independent and dependent
variables are upper limits, only Schmitt's binned method in ASURV is
available which may suffer from several drawbacks (Sadler et al.
1989). Hence we abandon upper limit points in $\alpha_{OX} - l_x$
plane and only use X-ray detected quasars to derive linear
regression parameters.
\end{enumerate}

Spearman correlation coefficients are calculated using ASURV
including X-ray undetected quasars. Observational data together with
linear regression slopes and Spearman correlation coefficients for
$\alpha_{OX} - l_o$, $l_x - l_o$ and $\alpha_{OX} - l_x$
correlations are shown in Figure 1. Conflicting correlations arise
due to dispersions in luminosity: $\alpha_{OX}=-0.18 l_o + const$ as
shown in solid line in Panel (a), but $\alpha_{OX}=0.20l_x  + const$
as shown in solid line Panel (c). Therefore the same data produce
two totally different results: $l_x=0.53 l_o + const$ or $l_x=2.08
l_o  + const$, i.e. $e<1$ or $e>1$ if $L_X \propto L_O^e$.
Therefore, depending upon how the regression is done, the conclusion
on the relationship between the optical and X-ray luminosities can
be significantly different. As shown in Panel (b), the slope of the
$l_x - l_o$ relation does depend on which luminosity is used as the
dependent variable. When treating $l_x$ as the dependent variable,
the slope is $0.54\pm0.14$ (the flatter dashed line), while it
changes dramatically to $2.08\pm0.45$ (the steeper dashed line) when
treating $l_o$ as the dependent variable. Using ILS bisector, we
find the slope to be $1.05\pm0.20$ (solid line), which is consistent
with $e=1$ (dotted line).

\begin{figure}
\includegraphics[width=84mm]{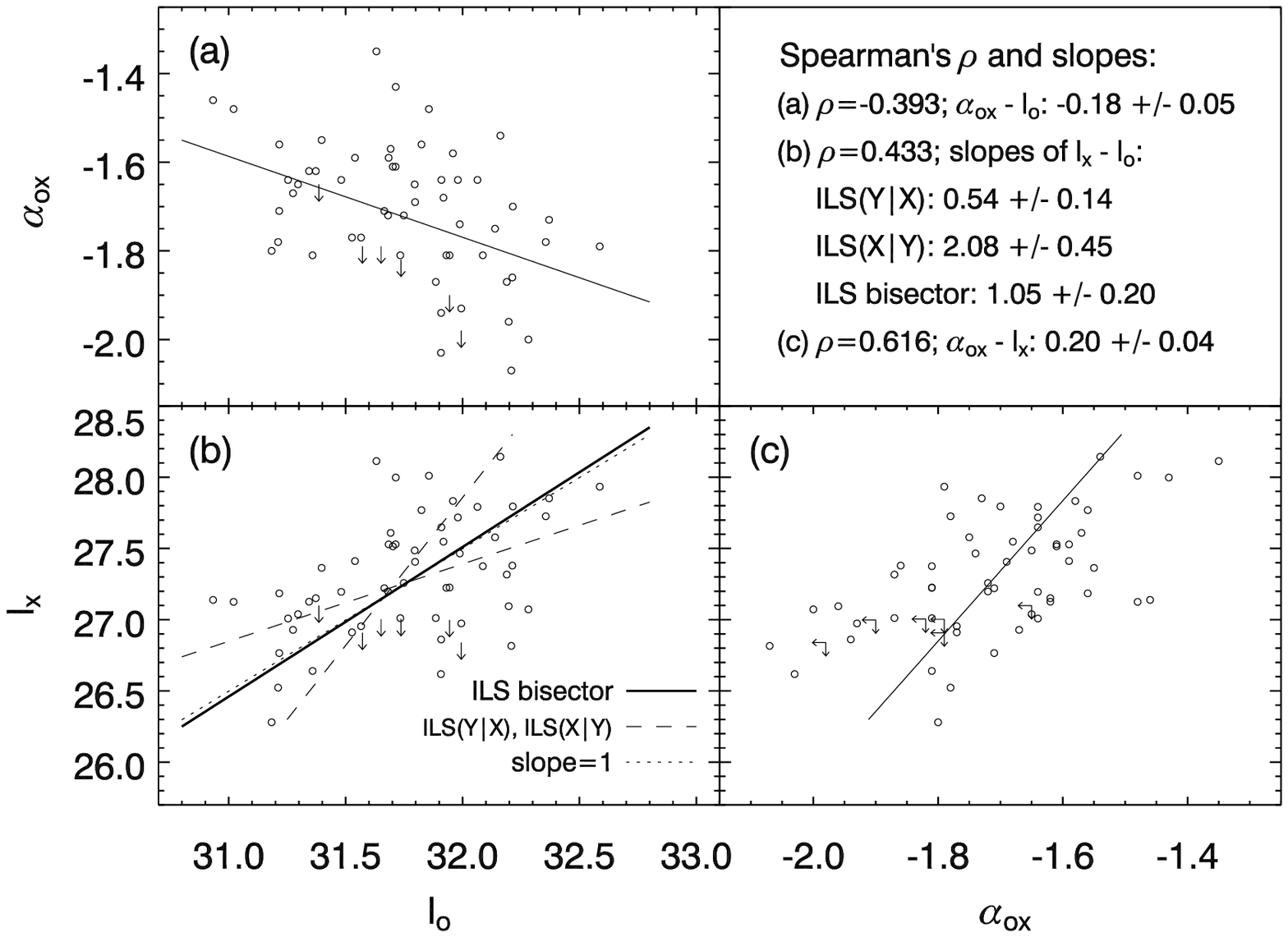}
\caption{61 high-redshift quasars in Miyaji et al. (2006). Circles
indicate X-ray detected quasars, while arrows indicate upper limits.
Panel (a) $\alpha_{OX}$ dependence on the rest frame 2500 $\rm \AA$
monochromatic luminosity. The solid line indicates linear regression
results from the EM Algorithm in ASURV. Panel (b) rest frame 2 keV
monochromatic luminosity against 2500 $\rm \AA$ one. The solid line
indicates the ILS bisector result, and dashed lines indicate
ILS(Y$\mid$X) and ILS(X$\mid$Y) results from the EM Algorithm in
ASURV, respectively. The $e=1$ relation is shown by dotted line for
comparison. Panel (c) $\alpha_{OX}$ dependence on the rest frame 2
keV monochromatic luminosity. The solid line indicates linear
regression result using X-ray detected quasars. Spearman correlation
coefficients and slopes of fitting lines are indicated in the upper
right panel.}
\end{figure}

The second part of analysis is done with Monte Carlo simulations.
Following Yuan et al. (1998), we assume intrinsic optical and X-ray
luminosities $\bar l_o$ and $\bar l_x$ with a constant mean $ \bar
\alpha_{OX}$,
\begin{equation}
\bar l_x= \bar l_o+\bar \alpha_{OX}/0.3838,
\end{equation}
and $\bar \alpha_{OX}=-1.728$ is given by the mean of observed
$0.3838(l_x-l_o)$ using the Kaplan-Meier estimator in ASURV,
including the six quasars with upper limits. The above relationship
is plotted as dotted line in Panel (b) of Figure 1. The observed
optical and X-ray luminosities are assumed to be the intrinsic
luminosities modified by independent Gaussian dispersions
\begin{equation}
l_o=\bar l_o + \delta l_o,  l_x=\bar l_x + \delta l_x= \bar l_o
+\bar \alpha_{OX}/0.3838 + \delta l_x,
\end{equation}
where $\delta l_o$ and $\delta l_x$ are Gaussian distributed
dispersions with standard deviations $\sigma_o$ and $\sigma_x$
respectively. Thus, the distribution of $\alpha_{OX}$ is Gaussian
with standard deviation
\begin{equation}
\sigma_{\alpha_{OX}}=0.3838(\sigma_o^2+\sigma_x^2)^{1/2},
\end{equation}
where $\sigma_{\alpha_{OX}}=0.154$ is the standard deviation around
the linear relationship of Equation (2) for this sample, using 55
X-ray detected AGNs. The ratio of the standard deviations of the
optical to the X-ray luminosity dispersion is defined as
\begin{equation}
R_{\sigma}=\frac{\sigma_o}{\sigma_x}.
\end{equation}

In the following we make Monte Carlo simulations by considering 21
values of $R_{\sigma}$ from 0.1 to 10, sampled evenly on a
logarithmic scale. We use the observed optical luminosity as $\bar
l_o$ and keep redshift unchanged. Then $l_o$, $l_x$ and fluxes can
be determined using Equations (3)-(5). The X-ray flux limit is
determined as follows. Quasars in this sample are from different
observations and thus not uniformly sampled. All quasars with $z\leq
3.5$ are X-ray detected. In the range of $z>3.5$, six quasars are
not X-ray detected. Five of them, i.e. SDSS 1737+5828, PSS
1435+3057, SDSS 1532-0039, PSS 1506+5220 and PSS 2344+0342
 were observed by {\it Chandra} with exposure times from 2.61-5.1
 ks; SDSS 0338+0021 was observed by {\it XMM-Newton} with exposure time 5.49 ks.
 PSS 1506+5220 has one count in 0.5-2 keV, and all the other five have zero counts.
 All the other detected sources have counts larger than 1. The average rest-frame
 $f_{2\ \rm keV}$ flux for one count in 0.5-2 keV of the six quasars is $0.8\times 10^{-32}$ erg cm$^{-2}$ s$^{-1}$
 Hz$^{-1}$. So we simply put a flux limit of $f_{2\ \rm keV,\
upper}=1.5\times0.8\times 10^{-32}$ erg cm$^{-2}$ s$^{-1}$
Hz$^{-1}$. We assume that simulated quasars with $z>3.5$ and $f_{2\
keV}$ less than $f_{2\ \rm keV,\ upper}$ will not be detected:
quasars with $0\leq f_{2\ \rm keV}<0.5f_{2\ \rm keV,\ upper}$ will
be assigned zero count, and quasars with $0.5f_{2\ \rm keV,\ upper}
\leq f_{2\ \rm keV}<1.5f_{2\ \rm keV,\ upper}$ will be assigned 1
count. Then the upper limits of non-detected quasars are at the 95\%
confidence level and will be calculated according to Kraft et al.
(1991), assuming one count corresponds to a flux of $0.8\times
10^{-32}$ erg cm$^{-2}$ s$^{-1}$.

Then we simulate 100 samples for each of 21 different $R_{\sigma}$
values. For each $R_{\sigma}$, we compute the average slopes and
Spearman correlation coefficients $\rho_{sp}$ from the 100 simulated
samples and display the results in Figures 2 and 3. Parameters for
simulated samples are calculated in the same way as for panels (a)
and (c) in Figure 1. As discussed in Strateva et al. (2005), the
measurement errors and variability effects bring $\sigma_x \sim
0.23$ and $\sigma_o \sim 0.17$, and $\sqrt{\sigma_{x}^2
+\sigma_{o}^2} \sim 0.29$. The observed dispersion is 0.4 for the
M06 sample and around $0.35-0.4$ for other samples. The extra
dispersions could be assigned to either $\sigma_x$ or $\sigma_o$ as
unknown dispersions, so the possible range of $R_{\sigma}$ should be
$0.5\sim1.4$, i.e. $\log(R_{\sigma}) \sim -0.3-0.15$. In the range
of $\log(R_{\sigma}) \sim -0.2-0$, $\rho_{sp}$ and slopes in
simulated samples for both $\alpha_{OX} - l_x$ and $\alpha_{OX} -
l_o$ correlations are consistent with observed values within
1.5$\sigma$, as shown in Figures 2 and 3.

\begin{figure}
\includegraphics[width=84mm]{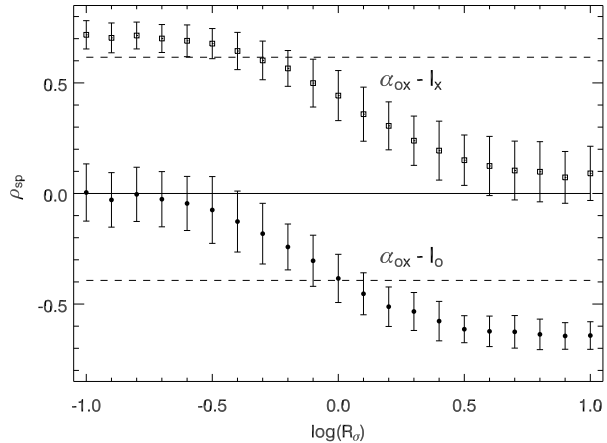} \caption{Spearman rank
correlation coefficients $\rho_{sp}$ as a measure of the
$\alpha_{OX}-l_o$ and $\alpha_{OX}-l_x$ correlation for Miyaji's
sample and simulated samples. Open squares: $\alpha_{OX}-l_x$ in
simulated samples; solid circles: $\alpha_{OX}-l_o$ in simulated
samples; upper dashed line: $\alpha_{OX}-l_x$ of true data; lower
dashed line: $\alpha_{OX}-l_o$ of true data; solid line indicates no
correlation.}
\end{figure}

\begin{figure}
\includegraphics[width=84mm]{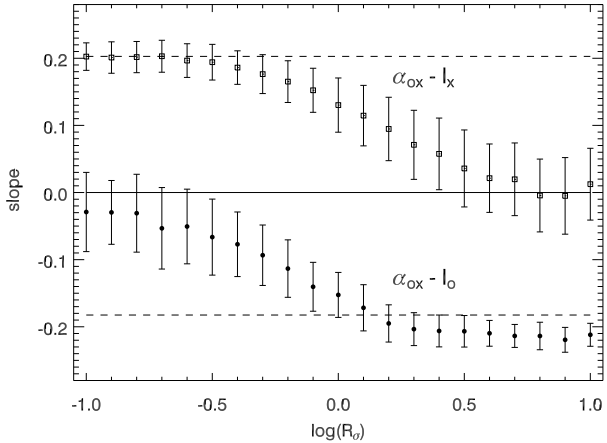} \caption{Slopes of the
$\alpha_{OX}-l_o$ and $\alpha_{OX}-l_x$ correlation for Miyaji's
sample and simulated samples. Symbols and lines are as in Figure
2.}
\end{figure}

From the above two analyses, we conclude that a constant
$\alpha_{OX}$ which does not depend on optical luminosity is still
consistent with data in this high-z sample.

\subsection{Determining factor for an artificial correlation caused by
luminosity dispersions}

Yuan et al. (1998) pointed out that a dispersion larger for the
optical luminosity than for the X-ray luminosity tends to result in
apparent, yet artificial correlation of $\alpha_{OX} - l_o$. To
quantitatively examine the effect of dispersions on the artificial
correlation, we make a simple analytic calculation as follows.

As shown in Figure 4, assuming $l_x=l_o+\rm const.$, dispersions in
$l_x$ and $l_o$ are $\sigma_x$ and $\sigma_o$, respectively, and the
span range of $l_o$ is $\Delta l_o$. Then locations of AGNs in the
$\alpha_{OX} - l_o$ plane with average $l_x$ and $1\sigma$ range of
$l_o$ are indicated by the dotted lines in the lower panel, where an
apparent correlation appears. We only consider the simplest
situation:
\begin{enumerate}
  \item AGNs are evenly distributed along $l_o$ (A concentration around a central $l_o$ is equivalent to
  a smaller $\Delta l_o$);
  \item AGNs are only distributed $0.3838f\sigma_o$ away from $\bar \alpha_{OX}$, i.e.
  $\alpha_{OX}= \bar \alpha_{OX} \pm 0.3838 f\sigma_o$, where $f$ is an unknown positive
  coefficient to be determined. We will discuss the value of $f$ later.
\end{enumerate}
Then we fit the observed AGNs in the $\alpha_{OX} - l_o$ plane,
assuming a least chi-squared fitting procedure with same weight in
$l_o$ and $\sigma_o$ (i.e. $\alpha_{OX}/0.3838$) as
\begin{equation}
\chi^2= [\sum_i ((l_o(i)-\hat l_o(i))^2+(\alpha_{OX}(i)-\hat
\alpha_{OX}(i))^2/0.3838^2)]/ \sigma_C^2,
\end{equation}
where the $l_o(i)$ and $\alpha_{OX}(i)$ are observed values, $\hat
l_o(i)$ and $\hat \alpha_{OX}(i)$ are the values in the fitting line
(indicated by a long solid line with a negative slope in the lower
panel of Figure 4) with the least $(l_o(i)-\hat
l_o(i))^2+(\alpha_{OX}(i)-\hat \alpha_{OX}(i))^2/0.3838^2$, and
$\sigma_C$ is the typical constant error of $l_o$. Then
\begin{equation}
\chi^2=\frac{k^2}{3(1+k^2)}(\frac{\Delta l_o^3}{4}+3\Delta l_o
(f\sigma_o)^2(1+\frac{1}{k})^2) / \sigma_C^2,
\end{equation}
where $k$ is the slope of the fitting line. The best fit slope $k$
can be derived by solving
\begin{equation}
\frac{d\chi^2}{d k}=0, \frac{d^2\chi^2}{d k^2}>0.
\end{equation}
The solution is
\begin{equation}
k_f=\frac{\Delta l_o^2}{24f^2\sigma_o^2}-\sqrt{(\frac{\Delta
l_o^2}{24f^2\sigma_o^2})^2+1} \\
\end{equation}
\begin{equation}
\simeq -12(\frac{f\sigma_o}{\Delta l_o})^2.
\end{equation}

The deviation of the approximation in Equation (10) from Equation (9) is less than
$20\%$ when $\frac{24 f^2\sigma_o^2}{\Delta l_o^2} < 1$. Considering two sub-samples
with the same weight and different slopes $k_1$ and $k_2$ respectively, the slope of
combined sample including all points in each sub-samples would be
$k_c=\tan((\arctan(k_1)+\arctan(k_2))/2) \simeq (k_1+k_2)/2$, where the last $\simeq$
is valid only if $k_1 \ll 1$ and $k_2 \ll 1$. Therefore, we can derive the slope of the
combined sample considering different $f$ values with different weights by an
integration
\begin{equation}
k=\int_{0}^{\infty} k_f p(f)df,
\end{equation}
where $p(f)$ is the probability of $f$.

Then we take the distribution of $f$ into account. If $\sigma_x=0$,
and $\sigma_o$ follows Gaussian distribution, the probability of $f$
would be
\begin{equation}
p(f)=\sqrt{\frac{2}{\pi}}e^{-\frac{f^2}{2}},\ f\geq0.
\end{equation}
Then
\begin{eqnarray}
k\simeq\int_{0}^{\infty} -12(\frac{f\sigma_o}{\Delta l_o})^2
\sqrt{\frac{2}{\pi}}e^{-\frac{f^2}{2}}df
=-6(\frac{\sigma_o}{\Delta l_o})^2 \\
\simeq \frac{\Delta l_o^2}{12\sigma_o^2}-\sqrt{(\frac{\Delta
l_o^2}{12\sigma_o^2})^2+1}.
\end{eqnarray}

Equation (13) shows that the slope of the artificial
$\alpha_{OX}-l_o$ correlation is directly proportional to
$\sigma_o^2/\Delta l_o^2$. The significance of the correlation,
which could be measured by Spearman correlation coefficient, is
always positively correlated to the slope value in the artificial
$\alpha_{OX}-l_o$ correlation, as shown in Figures 2-3. Therefore,
the result in Equation (13) also means the significance of the
artificial $\alpha_{OX}-l_o$ correlation is proportional to
$\sigma_o^2/\Delta l_o^2$.

For the M06 sample, $\Delta l_o \sim 1.7$ and $\sigma_o \sim 0.40$ when $\sigma_x=0$,
then $k\sim -0.3$. Such estimation of an artificial slope using Equations (13) or (14)
is qualitatively consistent with the slope in the $e=1$ simulations, where the value is
$k\sim -0.18$ as shown in Figure 3. A reason for the discrepancy is that the absolute
slope value is not $\ll 1$, therefore approximations used in our estimation are
deviated from true values. Another possible reason, i.e. different fitting procedure
used in simulations and our estimation, would more or less contribute to the lower
absolute slope values in simulations. In simulations, a linear regression method, which
only takes residuals of the dependent variable into account, is used, thus always
leading to a lower absolute slope value than methods considering residuals in both
variables as used in Equation (7), as shown in Panel (b) in Figure 1.

\begin{figure}
\includegraphics[width=84mm]{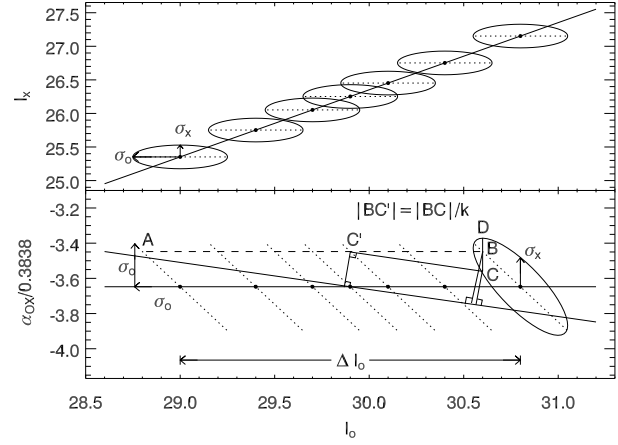} \caption{Schematic sketches for
the $l_x -l_o$ and $\alpha_{OX}-l_o$ relationship. Here $l_x=l_o +
\bar \alpha_{OX}/0.3838$ and $\bar \alpha_{OX}=-1.4$. See section
2.2 for details.}
\end{figure}

We now discuss consequences of non-zero $\sigma_x$. When
$\sigma_x>0$, the distribution of $f$ will be extended. As shown in
Figure 4, point B becomes a distribution in the range of CD within
$1\sigma$. When calculating the $\chi^2$ of a linear regression with
slope $-k$, the contribution of the broadening in B is equivalent to
an extension along the $l_o$ axis. For example, as shown in Figure
4, the contribution of point C is equivalent to point C' which has
the same $\alpha_{OX}$ as B but smaller $l_o$. Therefore, a non-zero
$\sigma_x$ tends to smooth the distribution of $l_o$ and extend its
range. When $\sigma_x/k$ is comparable with $l_o$, it will extend
the range of $l_o$ significantly. To show this effect, we do another
simulation. Based on the M06 sample with a given $\sigma_o=0.3$, we
calculate the Spearman correlation coefficients and slope for
$\alpha_{OX} - l_o$ correlation with different $\sigma_x$, while
other conditions are set to be the same as simulations in section
2.1. As shown in Figure 5, when $\log(\sigma_x/\sigma_o)>0$, i.e.
$\sigma_x>0.3$ and $\sigma_x/k > 1.7 \sim \Delta l_o$, both the
Spearman correlation coefficient and the slope tend to move toward
zero when $\sigma_x$ increases. When $\log(\sigma_x/\sigma_o)>0.5$,
i.e. $\sigma_x>0.9$ and $\sigma_x/k > 5.3 \sim 3\Delta l_o$, there
is no correlation in $\alpha_{OX} - l_o$ within $1\sigma$. However,
as discussed in Section 2.1, even if all extra dispersion in
$\alpha_{OX}$ comes from $l_x$, $\sigma_x$ is unlikely to exceed 0.4
and $\log(\sigma_x/\sigma_o)$ is unlikely to exceed $0.3$.

The effects of $\sigma_o$ on the relationship of $\alpha_{OX} - l_x$
is similar with the effects of $\sigma_x$ on the relationship of
$\alpha_{OX} - l_o$. Thus we do not need to repeat the above
analysis. Moreover, a similar effect as presented here for the
$\alpha_{OX} - l_o$ correlation would also affect any correlation
with a dependent variable $B$, which is not directly observed but
derived from $B\propto A^{-1}$, where $A$ is the independent
variable, such as the Baldwin effect, which has also been pointed
out by Yuan et al. (1998)

In summary, the significance of artificial correlation in $\alpha_{OX} - l_o$ is
approximately proportional to $\sigma_o^2/\Delta l_o^2$, and decreases when $\sigma_x$
increases and becomes comparable with $k\Delta l_o$, where $k$ is the absolute value of
the artificial slope and $\Delta l_o$ is the range $l_o$ span.

\begin{figure}
\includegraphics[width=84mm]{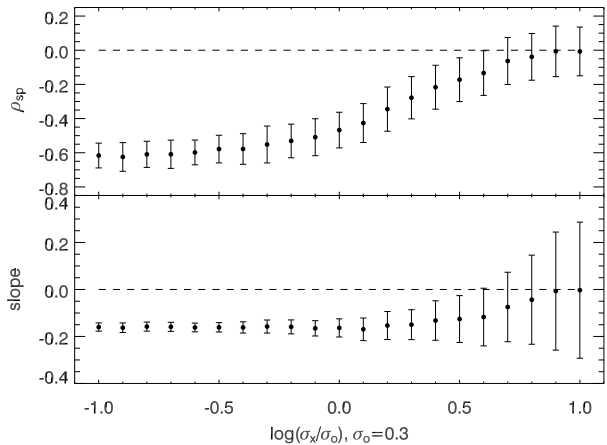} \caption{Spearman rank correlation
coefficients $\rho_{sp}$ and best-fit slopes of the
$\alpha_{OX}-l_o$ correlation for simulated samples for various
$\sigma_x$, with constant $\sigma_o=0.3$ and $\bar
\alpha_{OX}=-1.729$. $l_o$ and redshift distributions are from
Miyaji et al. (2006).}
\end{figure}

\section{Data Analysis of a sub-sample of AGNs from Steffen et al. (2006)}

Another problem in the study of $\alpha_{OX} - l_o$ relationship is
the degeneracy between redshift and luminosity in flux-limited
samples. S06 used a much larger sample than M06 with $\Delta l_o
\sim 5$, which suppresses the false slope artifacts discussed in
section 2. The observed change in $\alpha_{OX}$ across this larger
baseline $\Delta l_{o}$ in their sample is sufficiently large that
$\alpha_{OX}$ must depend on luminosity, or redshift, or both. To
distinguish between luminosity dependence and redshift dependence,
first, S06 performed partial correlation analysis using Kendall's
generalized partial $\tau$ to quantitatively show the correlation
significance of $\alpha_{OX} - l_o$ and $\alpha_{OX} - z$. They
found a 13.6$\sigma$ correlation of $\alpha_{OX} - l_o$ when
controlling $z$, and a 1.3$\sigma$ correlation of $\alpha_{OX} - z$
when controlling $l_o$. However, Kelley et al. (2007) has showed
that interpretation of Kendall's $\tau$ is problematic, and
Kendall's $\tau$ for the $\alpha_{OX} - z$ correlation is not
necessarily expected to be non-zero when $\alpha_{OX}$ is correlated
with $z$. Based on simulations, Kelley et al. (2007) pointed out
that the lack of evidence for a significant correlation between
$\alpha_{OX}$ and $z$ based on Kendall's $\tau$ in Steffen et al.
(2006) may be the result of an incorrect assumption about the
distribution of $\tau$ under the null hypothesis. In spite of most
previous studies, Kelley et al. (2007) found that $\alpha_{OX}$ is
correlated with both $l_o$ and $z$. Moreover, in the partial
correlation analysis of $\alpha_{OX} - l_o$, consequences of
luminosity dispersions, as discussed in section 2, were not taken
into account. To show this effect, we select AGNs in three redshift
bins, with each bin containing 38 sources, to control the redshift.
Then we examine the $\alpha_{OX} - l_o$ and $\alpha_{OX} - l_x$
relations in each bin, as shown in Figures 6 and 7. Similar to
Figure 1, in each redshift bin, $\alpha_{OX}$ is anti-correlated
with $l_o$, but positively correlated with $l_x$, which is caused by
luminosity dispersions. Because dispersions always strengthen the
anti-correlation of $\alpha_{OX} - l_o$, dependence of $\alpha_{OX}$
on $l_o$ might be biased toward higher significance by luminosity
dispersions, whereas the dependence on $z$ does not suffer such
bias.

Second, S06 compared $\alpha_{OX}$ residuals as a linear function of
$l_o$ and $z$. As shown in their Figure~8, there are systematic
residuals of $\alpha_{OX}-\alpha_{OX}(z)$, which indicate that
$\alpha_{OX}$ cannot be linearly dependent on redshift alone.
However, spectral index might depend on redshift in a non-linear
form, as shown in Figure 12 of Strateva et al. (2005). Moreover, it
is also possible that $\alpha_{OX}$ depends on both $l_o$ and $z$,
as shown in Kelley et al. (2007). Using different parameteric models
for the redshift and optical luminosity dependencies, Kelley et al.
(2007) found the model that is best supported by their data has a
linear dependence of $\alpha_{OX}$ on cosmic time, and a quadratic
dependence of $\alpha_{OX}$ on $l_o$ (the definition of
$\alpha_{OX}$ in Kelley et al. (2007) is different from our
definition with an opposite sign). Since $l_o$ and $z$ are coupled
together in flux limited samples, different parameteric models will
lead to different results and their best model results depend on the
form of the models. In summary, dependence of $\alpha_{OX}$ on $z$,
though with lower significance in partial correlation analysis,
could not be excluded in S06 sample.

\begin{figure}
\includegraphics[width=84mm]{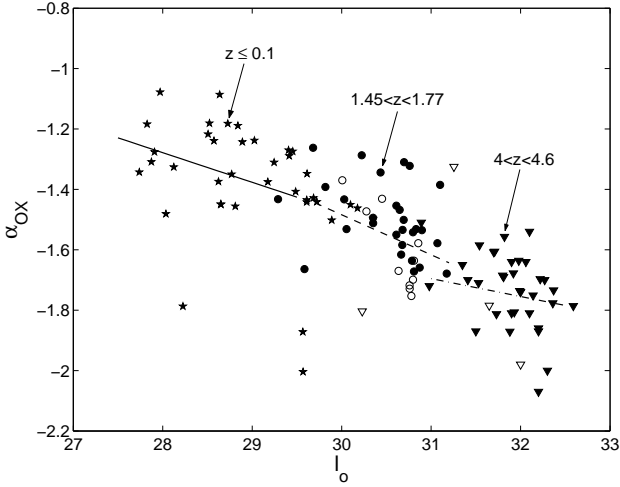} \caption{ $\alpha_{OX}$ dependece on the
 rest frame 2500 \AA\ monochromatic luminosity for AGNs in three redshift bins in S06 sample:
 $z\leq0.1$ (stars), $1.45<z<1.77$ (circles) and $4<z<4.6$ (downward-pointing
 triangles). X-ray detected AGNs are represented using filled
 symbols while upper limits are represented using open symbols. The
 solid line, dashed line and dash-dotted line indicate linear
 regression results from the EM Algorithm in ASURV for $z\leq0.1$, $1.45<z<1.77$ and
 $4<z<4.6$ samples, respectively.
 }
\end{figure}

\begin{figure}
\includegraphics[width=84mm]{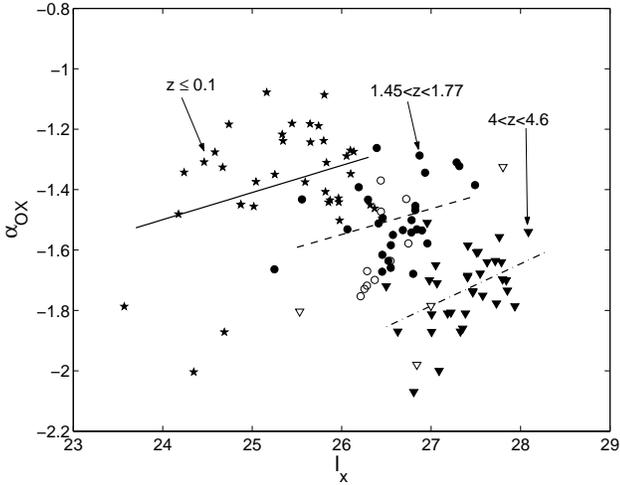} \caption{$\alpha_{OX}$ dependece on the
 rest frame 2 keV monochromatic luminosity for AGNs in three redshift bins in S06 sample:
 $z\leq0.1$ (stars), $1.45<z<1.77$ (circles) and $4<z<4.6$ (downward-pointing
 triangles). X-ray detected AGNs are represented using filled
 symbols while upper limits are represented using open symbols. The
 solid line, dashed line and dash-dotted line indicate linear
 regression results using X-ray detected sources for $z\leq0.1$, $1.45<z<1.77$ and
 $4<z<4.6$ samples, respectively.}
\end{figure}

To avoid possible bias from $\alpha_{OX} - z$ correlation, here we
examine a sub-sample from S06 containing 187 AGNs, as shown in the
dotted-line box in Figure 3 in S06 (Steffen et al. 2006; Strateva et
al. 2005; Vignali et al. 2005; Shemmer et al. 2005; Kelly et al.
2005), which more completely fills the redshift and optical
luminosity plane and thus is less affected by such bias. We refer
this sample as the `low-z sub-sample'. We perform linear regressions
in the same procedure as described in section 2.1. Figure 8 presents
our results for this sample. Similar to the M06 sample, the low-z
sub-sample show conflicting correlations due to dispersions in
luminosity: $\alpha_{OX}=-0.16l_o + const$ as shown in solid line in
Panel (a), but $\alpha_{OX}=0.11l_x  + const$ as shown in solid line
Panel (c). Therefore the same data produce two totally different
results: $l_x=0.58l_o  + const$ or $l_x=1.40l_o + const$, i.e. $e<1$
or $e>1$ if $L_X \propto L_O^e$. As shown in Panel (b), the slope of
the $l_x - l_o$ relation also depends on which luminosity is used as
the dependent variable. When treating $l_x$ as the dependent
variable, the slope is $0.59\pm0.06$ (the flatter dashed line),
while it changes dramatically to $1.45\pm0.12$ (the steeper dashed
line) when treating $l_o$ as the dependent variable. Using ILS
bisector, we find the slope to be $0.93\pm0.08$ (solid line), which
is consistent with $e=1$ (dotted line). In panel (c), the fit looks
different from the trend by eye (bisector fit). As discussed in
section 2, when using traditional ordinary least-squared method
which minimizes the residuals of the dependent variable, the fit
tends to be flatter than the bisector one where residuals of both
dependent and independent variables are taken into account. When
data points are concentrated at the center, as in Figure 8(c),
inconsistence of the two fits becomes large.

\begin{figure}
\includegraphics[width=84mm]{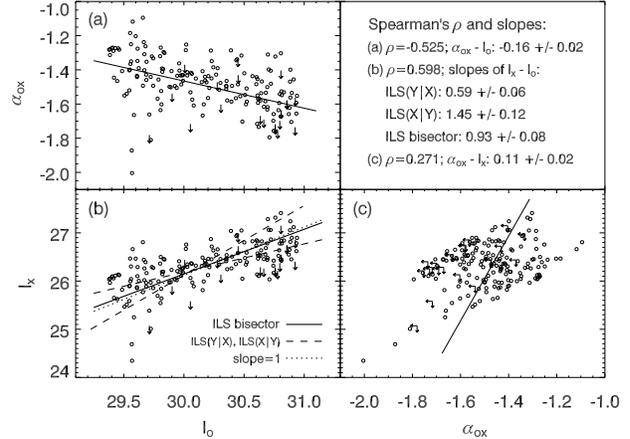} \caption{187 low-redshift AGNs from Steffen et al. (2006).
 Circles indicate X-ray detected AGNs, while arrows indicate upper limits.
Panel (a) $\alpha_{OX}$ dependence on the rest frame 2500 $\rm \AA$
monochromatic luminosity. The solid line indicates linear regression
results from the EM Algorithm in ASURV. Panel (b) rest frame 2 keV
monochromatic luminosity against 2500 $\rm \AA$ one. The solid line
indicates the ILS bisector result, and dashed lines indicate
ILS(Y$\mid$X) and ILS(X$\mid$Y) results from the EM Algorithm in
ASURV, respectively. The $e=1$ relation is shown by dotted line for
comparison. Panel (c) $\alpha_{OX}$ dependence on the rest frame 2
keV monochromatic luminosity. The solid line indicates linear
regression result using X-ray detected quasars. Spearman correlation
coefficients and slopes of fitting lines are indicated in the upper
right panel.}
\end{figure}

Interestingly, but not surprisingly, results of ILS$(\alpha_{OX}\mid
l_o)$ are consistent with results of ILS$(l_x \mid l_o)$, and
results of ILS$(\alpha_{OX}\mid l_x)$ are consistent with results of
ILS$(l_o \mid l_x)$, as shown in Table 1 and Figures 1 and 8. The
reason is that artificial correlations seen in $\alpha_{OX} - l_o$
relation, i.e. $0.3838(l_x-l_o) - l_o$ relation, corresponds to
least-squares method which minimizes the residuals of $l_x$, which
is the same as in $l_x - l_o$ relation, and thus leads to a slope
less than the true one where residuals of both variables are
considered. Such results indicate that for $l_x - l_o$ relation,
regression results based on the traditional ordinary least-squares
method which minimizes residuals of the dependent variable suffer
from effects caused by luminosity dispersion and are thus not
reliable. Instead, weighting both $l_o$ and $l_x$ in the regression,
i.e. ILS bisector, would be a more robust methodology. Moreover, the
degree to which the slopes of ILS$(\alpha_{OX}\mid l_x)$ and
ILS$(\alpha_{OX}\mid l_o)$ are inconsistent indicates the degree of
artificial correlation discussed in section 2. When the slopes
converge to the same value, the artificial correlation would be
suppressed.

\begin{table}
 \caption{Slopes of $l_x - l_o$ relation in M06 and low-z sub-sample derived from different regression methods.}
 \label{symbols}
 \begin{tabular}{lcc}
  \hline
  Regression method & M06 &  low-z sub-sample \\
  \hline
  ILS$(\alpha_{OX}\mid l_o)$ &  0.53 $\pm$ 0.13 & 0.58 $\pm$ 0.05 \\
  ILS$(l_x \mid l_o)$ & 0.54 $\pm$ 0.14 & 0.59 $\pm$ 0.06 \\
  \hline
  ILS$(\alpha_{OX}\mid l_x)$ & 2.08 $\pm$ 0.48 & 1.40 $\pm$ 0.10 \\
  ILS$(l_o \mid l_x)$ & 2.08 $\pm$ 0.45 & 1.45 $\pm$ 0.12\\
  \hline
 \end{tabular}
\end{table}

We conclude for this sample that a constant $\alpha_{OX}$ which does
not depend on luminosity is also consistent with data.

\section{Comparison of Optical and X-ray Luminosity Functions}
For a complete sample of broad line AGN including optical and X-ray
observations (detections or upper limits), slopes in optical and
X-ray quasar luminosity functions (LFs) should be the same after
correct transformations. When converting optical luminosity to X-ray
luminosity, different $\alpha_{OX}$ models would lead to different
X-ray LF shapes in the optical frame. Thus we can test whether a
particular $\alpha_{OX}$ model is correct by comparing the two LFs.
We use the optical quasar LF from Richards et al. (2006) and AGN
hard X-ray LF from Barger et al. (2005).

We investigate the following two $\alpha_{OX}$ models:
\begin{enumerate}
  \item a constant $\alpha_{OX}$, $l_x=l_o-3.92$;
  \item $\alpha_{OX}$ from Steffen et al. (2006),
$l_x=0.72l_o+4.53$;
\end{enumerate}

We follow Hopkins, Richards \& Hernquist (2007) to calculate the
binned LFs. For each $\alpha_{OX}$ model, an overall normalization
factor is applied in the X-ray LF to get the minimum $\chi^2$, which
means that we are comparing just the slopes of LFs. Results are
presented in Figure 9. A constant $\alpha_{OX}$ which does not
depend on luminosity (left panels) is consistent with data, and is
more preferred than the $\alpha_{OX}$ models given by S06 (right
panels).

However, from this comparison we cannot reach a strong conclusion
that luminosity dependent $\alpha_{OX}$ is excluded completely, for
three reasons as follows. First, there are very few data points here
in the X-ray LF. Second, as pointed out by Richards et al. (2005),
such comparison is not strictly quantitative since X-ray selected
samples and optically selected samples are not identical. Moreover,
the bright-end slopes from different X-ray samples are different
(Barger et al. 2005; Ueda et al. 2003; Hasinger, Miyaji, \& Schmidt
2005). Hopkins, Richards \& Hernquist (2007) combined a large set of
LF measurements and took obscuration and scattering into account,
and in their analysis the luminosity functions can be reconciled
reasonably well with the $\alpha_{OX}$ model in S06. However, since
the constraints on the present bright-end X-ray LFs are poor, the
fact that the LFs could be reconciled reasonably well with S06 in
their work probably just reflects the large X-ray error bars.

\begin{figure}
\includegraphics[width=84mm]{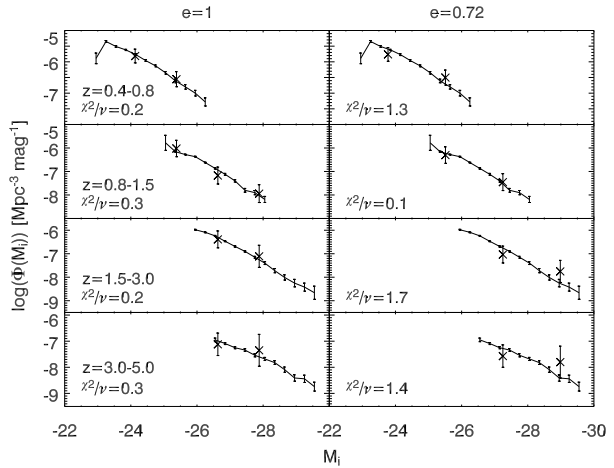} \caption{Comparison of the SDSS DR3 optical quasar
luminosity function from Richards et al. (2006) (lines with
errorbars) with AGN hard X-ray luminosity function from Barger et
al. (2005) (crosses with errorbars). Left panels show the LFs with
$e=1$ (luminosity independent spectral index). Right panels show the
results with $e=0.72$ (spectral index is anti-correlated with
luminosity). Different rows show results at different redshift
range. For each panel, an overall normalization factor is applied in
X-ray LF to get the minimum $\chi^2$. }
\end{figure}

\section{Selection effects in flux limited samples: optically selected samples vs X-ray selected samples}
In this section, we discuss the selection effects in flux limited
samples.  For a given AGN luminosity function, assuming the observed
optical and X-ray luminosities are the intrinsic values modified by
dispersions which might be caused by variabilities or observational
errors, there are three possibilities in a flux limited sample:

a) lower fraction of more luminous AGNs are missed;

b) higher fraction of more luminous AGNs are missed;

c) same fractions of more luminous and fainter AGNs are missed, so
the relationship between $l_x - l_o$ remains unchanged.

Assuming the slope of $l_x - l_o$  relation (without further description, slope=$e$ in
$l_x=e\times l_o + const$ throughout this section) is unity, and at a certain redshift
the number density of AGN decreases with luminosity, Figure~10 shows schematic sketches
for the first two cases in optically selected AGN samples. The upper panels are for
case (a), where the density contour lines in luminosity functions in the $l_o - z$
plane (each line corresponds to a given constant AGN number density as a function of
redshift) are steeper than the flux limits, hence lower fraction of more optical
luminous AGNs are missed, and then the slope is biased toward more than unity. The
bottom panels are for case (b), where the density contour lines in luminosity functions
in the $l_o - z$ plane are flatter than the flux limits, hence higher fraction of more
optical luminous AGNs are missed, and then the slope is biased toward less than unity.
The left panels show flux limits (dashed lines) in $l_o-z$ plane, compared with the
density contour lines in luminosity functions (solid lines). The left panels show
observed AGNs (solid circles) which are above the flux limits (dashed lines), and
missed AGNs (open circles) which are below the flux limits, in $l_o - l_x$ plane.
Slope$=1$ are indicated by solid lines.

\begin{figure}
\includegraphics[width=84mm]{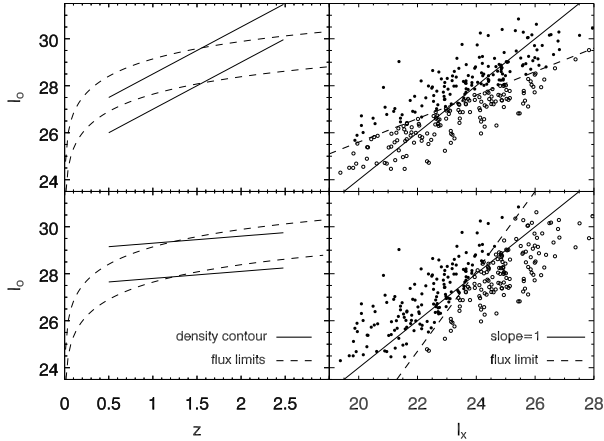} \caption{Schematic sketches for
selection effects in optically selected AGN samples. The upper panels are for case (a)
and the bottom panels are for case (b). The left panels show flux limits (dashed lines)
in $l_o-z$ plane, compared with the density contour lines in luminosity functions
(solid lines). The left panels show observed AGNs (solid circles) which are above the
flux limits (dashed lines), and missed AGNs (open circles) which are below the flux
limits, in $l_o - l_x$ plane. Slope is defined as in $l_x=$ slope$\times l_o + const$.
Slope$=1$ are indicated by solid lines. Note that in the upper-right panel slope$>1$
for detected AGNs, and in the lower-right panel slope$<1$ for detected AGNs, since
$l_x$ is the x-axis and $l_o$ is the y-axis. These sketches, much simplified and only
qualitatively correct, are shown for illustration only.}
\end{figure}

The density contour lines in real luminosity functions are much more
complicated than shown in Figure~10, and the real slopes of density
contour lines depend on redshifts and luminosities. Moreover, since
the fraction of missed AGNs depends on the dispersions of
luminosities around the linear relationship of $l_x - l_o$, the
biases also depends on the dispersions. To test whether the slope of
$l_x - l_o$ relation could be biased by flux limits in realistic
optically selected AGN samples, we carry out Monte Carlo
simulations. We use optical analytical luminosity function from
Richards et al. (2005) for $z<3$ AGNs, and Richards et al. (2006)
for $z>3$ AGNs. We simulate three optically selected sub-samples, in
order to mimic the SDSS, COMBO-17 and high-z samples in S06:

1) a shallow $z<3$ sub-sample with $m_g<19$ containing about 155
AGNs, which is similar to the SDSS sample in S06;

2) a deeper $z<3$ sub-sample with $m_g<21$ containing about 52 AGNs,
which is similar to the COMBO-17 sample in S06;

3) a $4<z<6$ sub-sample with $m_g<20$ containing about 55 AGNs,
which is similar to the high-z sample in S06.

We do not try to mimic the nearby Seyfert 1 and BQS samples in S06 which are located in
$z<0.4$, since the LF in this redshift range has larger errors due to smaller volume.
For the two $z<3$ sub-samples, a detection efficiency factor is taken from Figure 6 in
Richards et al. (2006). For the $z>4$ sub-sample, a constant detection efficiency is
used according to Richards et al. (2006). To show the effects of luminosity dispersions
in optically selected flux limited samples, assuming the slope of $l_x - l_o$ equals
unity with dispersions, 1000 simulations, each containing the above three sub-samples,
are carried out for each of the following five dispersion models:
\begin{enumerate}
  \item $\sigma_o=0.42$, $\sigma_x=0$;
  \item $\sigma_o=\sigma_x=0.3$ for $z<3$ AGNs, $\sigma_o=\sigma_x=0.5$ for $z>4$ AGNs;
  \item $\sigma_o=\sigma_x=0.3$
  \item $\sigma_o=\sigma_x=0.5$
  \item $\sigma_o=0$, $\sigma_x=0.42$.
\end{enumerate}
where $\sigma_o$ and $\sigma_x$ are defined as in section 2.1.
Models (i) and (v) are extreme cases, in order to show the bias
direction when $\sigma_o$ or $\sigma_x$ is dominating. Models (ii)
$\sim$ (iv) can show the effects of dispersion, and dispersion
evolution in cosmic time.

The Monte Carlo analysis was performed by generating a combined sample containing the
above three sub-samples for each of the five dispersion models as follows: first, the
redshift and luminosity ranges are divided into grids with $\delta z=0.1$ and $\delta
M_g = 0.3$, then the detection probability in a given grid is proportional to $d
V(z)\times \Phi(z, M_g) \times \eta$, where $d V(z)$ is the volume element in comoving
space, $\Phi(z, M_g)$ is the luminosity function, and $\eta$ is the detection
efficiency. In a given grid, AGNs are randomly produced following an uniform
distribution, and the total number of AGNs in the grid is determined by a poisson
process with expectation $\bar{N}(z, M_g)=d V(z)\times \Phi(z, M_g) \times \eta \times
C$, where C is a constant for a given sub-sample with given luminosity dispersions,
adjusted to make the average number of detected AGNs to be 155, 52 and 55 for the three
sub-samples, respectively. The conversion from $M_g$ to the intrinsic optical
luminosity $\bar{l}_o$ follows Richards et al. (2005), and the intrinsic X-ray
luminosity $\bar{l}_x=\bar{l}_o-4$. Second, luminosity dispersions are applied, where
the observed $l_o$ and $l_x$ are drawn from the Gaussian distribution around
$\bar{l}_o$ and $\bar{l}_x$ with given dispersions $\sigma_o$ and $\sigma_x$,
respectively. Third, flux limits are applied, AGNs with $m_g >$ flux limit are
detected. The above procedure select about 262 AGNs for each dispersion model, where
about 155 in sub-sample 1), 52 in sub-sample 2) and 55 in sub-sample 3). Then the above
procedure is repeated 1000 times to get 1000 independent samples.

The slope in each simulated sample is calculated using FITEXY (Press et al. 1992)
assuming the same error in $l_o$ and $l_x$. The distributions of slopes in simulated
optically selected flux limited samples are shown in Figure~11. The relative
probabilities are normalized with peak values equal unity. From left to right are the
five dispersion models (i) to (v) respectively. The mean values $\pm$ standard
deviations of slopes in simulations of the five models are: $0.76\pm0.05,\
0.89\pm0.06,\ 0.97\pm 0.05,\ 1.04\pm 0.11,$ and $1.23\pm0.06$, respectively. In models
(iii) and (iv), $\sigma_o=\sigma_x=0.3$ and 0.5, flux limited simulations are
consistent with the assumed slope$=1$, i.e. the slopes are not biased. However, when
$\sigma_o \neq \sigma_x$ or $\sigma_o,\ \sigma_x$ change in cosmic time, slopes in flux
limited simulations might be biased, as shown in models (i), (ii) and (v). Slopes in
model (i), i.e. $0.76\pm0.05$, is consistent with the slope in S06, i.e. $0.72\pm
0.01$. However, since $\sigma_o=0.42$ and $\sigma_x=0$ is an unrealistic extreme case,
this does not mean that the non-unity slope of $l_x - l_o$ relation is totally caused
by such selection effects. Note that slope values in Figures~11 depend on the selection
method, i.e. flux limits and number of AGNs in each sub-sample, hence another different
optical samples will have different results.

\begin{figure}
\includegraphics[width=84mm]{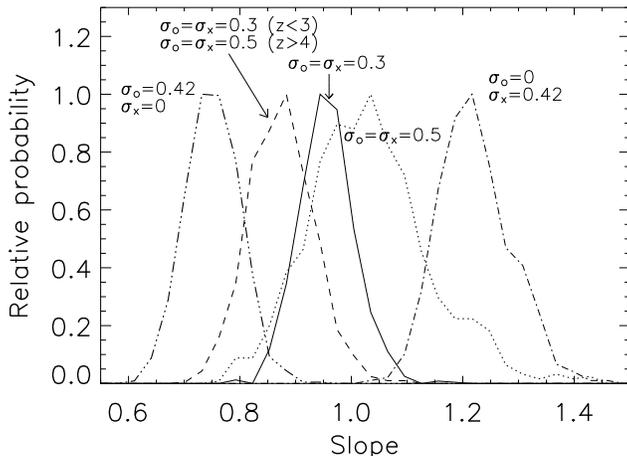} \caption{The distributions of slopes in simulated optically selected flux
limited samples, where $l_x=$slope$\times l_o + $constant. The
relative probabilities are normalized with peak values equal unity.
>From left to right are the five dispersion models (i) to (v) listed
in the text, respectively.}
\end{figure}

We do not simulate X-ray selected flux limited samples for two reasons. First, if the
slope of $l_o - l_x$ equals unity, X-ray AGNs are identical to optical AGNs, and the
X-ray LF will be the same as optical LF with $l_x=l_o+$ constant, as suggested in
section 4. Therefore, the results here can be applied to X-ray selected sample with
similar flux limits after switching $\sigma_o$ and $\sigma_x$. Second, X-ray LFs have
larger errors than optical LFs due to smaller samples. Therefore, it is our purpose to
just point out the fact that the slope will be biased in flux limited X-ray samples,
rather than focusing on a particular X-ray selected sample.

While a number of previous studies of optical selected AGNs have
reported that $\alpha_{OX}$ is anti-correlated with luminosity (e.g.
Strateva et al. 2005; Steffen et al. 2006), whether $\alpha_{OX}$
depends on luminosity in X-ray selected AGN samples remains unknown.
Hasinger (2004) found no $\alpha_{OX}$ dependence on either
luminosity or redshift in soft X-ray selected samples. Frank et al.
(2007) found in their Chandra Deep Field-North sample
$l_x=(0.808\pm0.047)l_o+const.$. They also found the slope decreases
when only brighter sources are included, and the slope increases
when only fainter sources are included. It is possible that the
slope might be biased in this flux limited sample and the magnitude
of biases are different when using different flux limit. As shown in
Figure~11, if $\sigma_o> \sigma_x$, an optically selected sample
like in S06 will be biased toward a flatter slope. Moreover, if the
X-ray LF is similar to optical LF and the X-ray sample is consisted
of AGNs with similar flux limits, the X-ray sample will be biased
toward a steeper slope. Therefore, even if optically selected AGNs
and X-ray selected AGNs are identical, the slope in the optically
selected sample will be flatter than the slope in the X-ray selected
sample, which can properly explain the observed discrepancy.

In summary, selection effects in flux limited samples might bias the
$l_o - l_x$ relation and cause discrepancy in the $l_o - l_x$
relation in optically selected samples and X-ray selected samples,
especially when $\sigma_o \neq \sigma_x$ or $\sigma_o,\ \sigma_x$
change in cosmic time. The magnitude of the bias and discrepancy
depend on the luminosity function, flux limits of the sample, and
dispersions in optical and X-ray luminosities. Note that even if
such selection effects do bias the slope of the $l_o - l_x$ relation
toward the observed discrepancy between optical and X-ray samples,
it is not necessarily the only reason. It is possible that optically
selected samples and X-ray selected samples are consisted of
different AGNs, so slopes of the $l_o - l_x$ relation in optically
selected samples will be different from slopes in X-ray selected
ones. As discussed in Brusa et al. (2007), about $40\%$ of the X-ray
selected AGNs in their COSMOS sample would have not been easily
selected as AGN candidates on the basis of purely optical criteria,
either because similar colors to dwarf stars or field galaxies, or
because they are not point like sources in morphological
classification. Moreover, optically selected AGNs and X-ray selected
AGNs might be typically in different evolution stages and thus are
not identical (Shen et al. 2007).

\section{Discussion and Conclusions}
In summary, we have investigated the correlation between the
spectral index $\alpha_{OX}$ and optical/X-ray luminosities in AGNs
by means of linear regressions, Monte Carlo simulations, simplified
analytic estimations and comparison of X-ray and optical luminosity
functions. We have reached five conclusions:

1. The dependence of $\alpha_{OX}$ on optical luminosity found in
Miyaji et al. (2006) may not be an underlying physical property. It
remains unknown whether $e<1$ or $e>1$ if $L_X \propto L_O^e$ in
this high-z sample.

2. The luminosity dependence can be artificially generated very
easily by luminosity dispersions. The significance of artificial
correlation in $\alpha_{OX} - l_o$ is approximately proportional to
$\sigma_o^2/\Delta l_o^2$, where $\sigma_o$ is the optical
luminosity dispersion and $\Delta l_o$ is the range that $l_o$
spans, and decreases when $\sigma_x$ increases and becomes
comparable with $k\Delta l_o$, where $k$ is the absolute value of
the artificial slope. This effect also affects the Baldwin effect.
Instead of regressions only weighting one variable, weighting both
$l_o$ and $l_x$, i.e. ILS bisector, in the regression would be a
more robust methodology to avoid such bias.

3. In a more complete low-z sub-sample from Steffen et al. (2006),
$\alpha_{OX}$ must depend on luminosity, or redshift, or both.
However, a luminosity independent $\alpha_{OX}$ is still consistent
with data. Redshift dependencies cannot be ruled out and may be
large, but somewhat hidden because of luminosity dispersions, which
generate artificial luminosity correlations in each redshift bin.

4. In the comparison of X-ray (Barger et al. 2005) and optical
quasar (Richards et al. 2006) LFs, a luminosity independent
$\alpha_{OX}$ is consistent with data, and more preferred than the
luminosity dependent $\alpha_{OX}$ model given by S06.

5. Selection effects in flux limited samples might bias the $l_o -
l_x$ relation and cause discrepancy in the $l_o - l_x$ relation in
optically selected sample and X-ray selected sample, especially when
$\sigma_o \neq \sigma_x$ or $\sigma_o,\ \sigma_x$ change in cosmic
time. The magnitude of the bias depends on the luminosity function,
flux limits of the sample, and dispersions in optical and X-ray
luminosities.

It therefore remains inconclusive whether the anti-correlation
between AGN spectral index and optical luminosity is true. Even if
$\alpha_{OX}$ does depend on optical luminosity, the currently
adopted slope value might be biased and deviate from the intrinsic
value. To correctly establish a dependence of $\alpha_{OX}$ of AGNs
on their luminosity, a larger and more complete sample, such as from
multi-wavelength surveys, is needed and consequences of luminosity
dispersions and selection effects in flux limited samples must be
taken into account properly.

\section*{Acknowledgments}

We thank the anonymous referee for helpful comments and stimulating
suggestions that improved the manuscript. S. M. T. thanks J. X.
Wang, G. T. Richards, N. Brandt and X. L. Zhou for helpful
discussion. S. N. Z. acknowledges partial funding support by the
Ministry of Education of China, Directional Research Project of the
Chinese Academy of Sciences and by the National Natural Science
Foundation of China under project no. 10233010 and 10521001.

\bsp

\label{lastpage}

\end{document}